\documentclass{elsart4-1}

\usepackage{graphicx,subfigure}

\usepackage{amssymb}

\usepackage[english,francais]{babel}
\usepackage[dvips]{color}

\newtheorem{e-proposition}[theorem]{Proposition}

\newtheorem{e-definition}[theorem]{Definition\rm}

\renewcommand{\theequation}{\arabic{equation}}
\newcommand{\tOmega}{\hat \Omega}
\newcommand{\bin}[2]{\left (^{#1}_{#2}\right )}
\newcommand{\coef}{{\mathcal S}}

\def\og{\leavevmode\raise.3ex\hbox{$\scriptscriptstyle\langle\!\langle$~}}
\def\fg{\leavevmode\raise.3ex\hbox{~$\!\scriptscriptstyle\,\rangle\!\rangle$}}

\begin{document}

\centerline{Physics} 
\begin{frontmatter}


\selectlanguage{english}
\title{Organic conductors in high magnetic fields:\\ model systems for quantum oscillations physics}


\selectlanguage{english}

\author[authorlabel2]{Alain Audouard},
\ead{alain.audouard@lncmi.cnrs.fr}
\author[authorlabel1]{Jean-Yves Fortin}
\ead{fortin@ijl.nancy-universite.fr}

\address[authorlabel2]{Laboratoire National des Champs Magn\'{e}tiques
Intenses (UPR 3228 CNRS, INSA, UJF, UPS) 143 avenue de Rangueil,
F-31400 Toulouse, France.}
\address[authorlabel1]{Institut Jean Lamour, Groupe de Physique Statistique,
CNRS-UMR 7198- Nancy-Universit\'e
BP 70239 F-54506 Vandoeuvre les Nancy Cedex, France}


\medskip
\begin{center}
{\small Received \today; accepted after revision +++++}
\end{center}

\begin{abstract}
Even though organic conductors have complicated crystalline structure with low
symmetry and large unit cell, band structure calculations predict multiband quasi-two dimensional electronic structure yielding very simple Fermi surface in most cases. Although few puzzling experimental results are observed, data of numerous compounds are in agreement with calculations which make them
suitable systems for studying magnetic quantum oscillations in networks of orbits connected by magnetic breakdown. The state of the art of this problematics is reviewed.
{\it To cite this article: A. Audouard, J-Y. Fortin, C. R.
Physique (2012).}

\vskip 0.5\baselineskip

\selectlanguage{francais}
\noindent{\bf R\'esum\'e}
\vskip 0.5\baselineskip
\noindent
{\bf Les conducteurs organiques sous champs magn\'{e}tiques intenses~: syst\`{e}mes mod\`{e}les pour la physique des oscillations quantiques. }
Bien que  les conducteurs organiques pr\'{e}sentent des structures cristallines
complexes et de basse sym\'{e}trie, les calculs pr\'{e}disent une structure \'{e}lectronique multibande quasi-bi dimensionnelle conduisant g\'{e}n\'{e}ralement \`{a} une surface de Fermi tr\`{e}s simple. Bien que quelques r\'{e}sultats exp\'{e}rimentaux d\'{e}routants soient observ\'{e}s, les donn\'{e}es de nombreux compos\'{e}s sont en accord avec les calculs, ce qui fait de ces derniers des syst\`{e}mes mod\`{e}les pour la physique
des oscillations quantiques dans les r\'{e}seaux d'orbites coupl\'{e}es par la rupture magn\'etique. L'\'{e}tat de l'art de cette probl\'{e}matique est pass\'{e}e en revue.
{\it Pour citer cet article~:  A. Audouard, J-Y. Fortin, C. R.
Physique (2012).}

 \vskip 0.5\baselineskip
\noindent{\small{\it Mots-cl\'es~:} M\'{e}taux organiques~; Oscillations quantiques~;
Surface de Fermi}}
\end{abstract}
\end{frontmatter}

\selectlanguage{francais}

\selectlanguage{english}
\section{Introduction}
\label{}

\begin{figure}                                                    
\centering
\includegraphics[width=0.9\columnwidth,clip,angle=0]{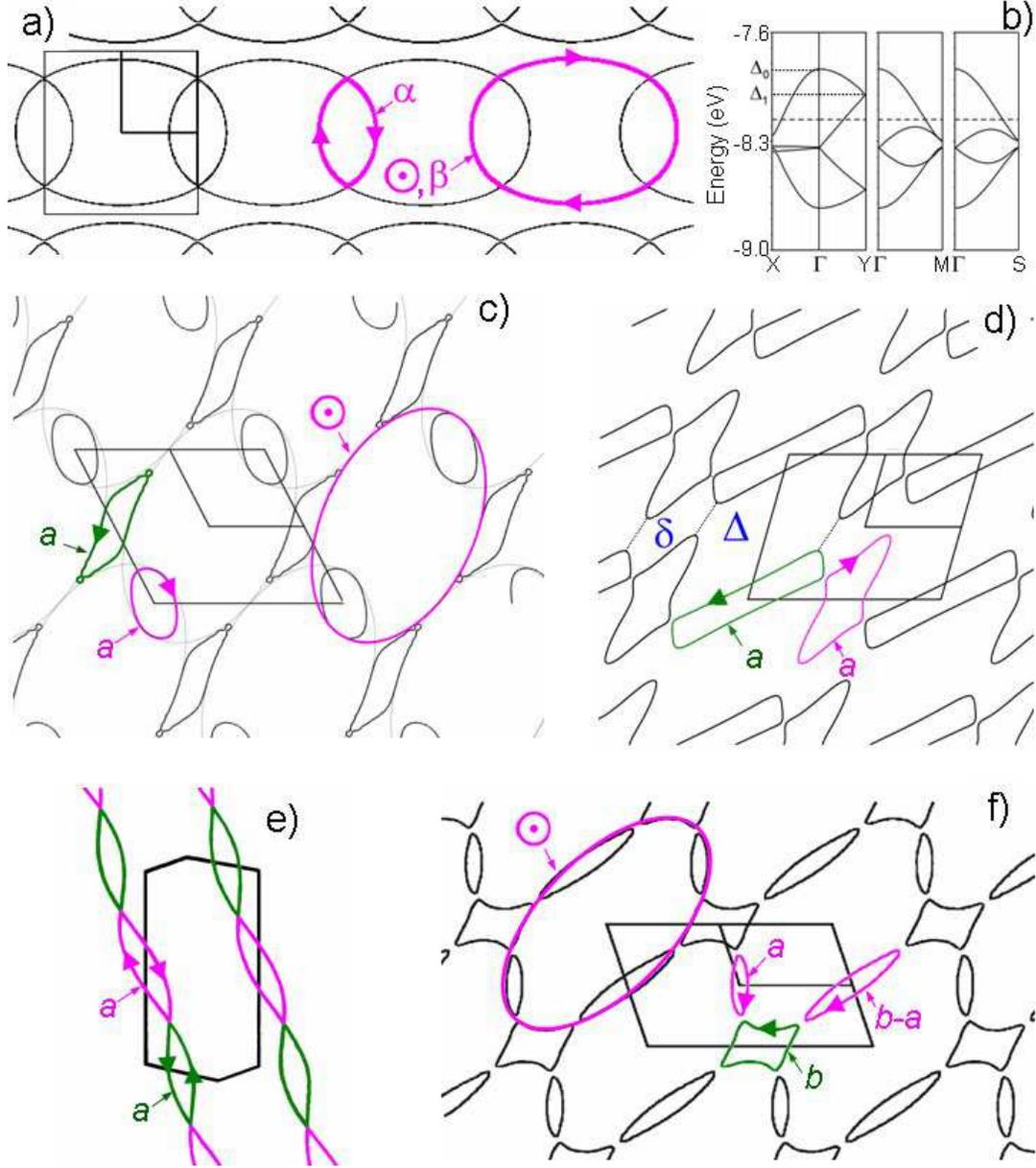}
\caption{\label{Fig:SF_networks}(Colour on-line)   Fermi surface (FS) of (a)
$\theta$-(ET)$_4$CoBr$_4$(C$_6$H$_4$Cl$_2$) \cite{Au12}, (c)
$\beta$''-(ET)$_4$NH$_4$[Fe(C$_2$O$_4$)$_3$]$\cdot$(C$_3$H$_7$NO) \cite{Pr03},
(d) (ET)$_8$[Hg$_4$Cl$_{12}$(C$_6$H$_5$Br)$_2$] \cite{Vi08}, (e)
(TMTSF)$_2$NO$_3$, in the temperature range between the anion ordering and the
spin density wave condensation \cite{Kan95} and (f)
(BEDO)$_4$Ni(CN)$_4\cdot$4CH$_3$CN \cite{Du05}. These FS achieve networks of
hole-type (pink solid lines) and electron-type (green solid lines) orbits. Arrows
indicate the quasi-particles path in magnetic field. Labels $\alpha$, $\beta$,
$a$, $b$, $b-a$ and $\bigodot$ stand for few classical orbits. $\delta$ and
$\Delta$ are FS pieces corresponding to forbidden orbits. (b) Energy dispersion of $\theta$-(ET)$_4$CoBr$_4$(C$_6$H$_4$Cl$_2$) \cite{Au12}.}
\end{figure}

 Quasi-two dimensional (q-2D) organic metals have generally complicated
crystalline structures with unit cell involving hundreds of atoms and cell
parameters as large as few dozens of nm. Nevertheless, band structure
calculations predict very simple Fermi surface (FS) which allow to view these
compounds as model systems for quantum oscillations physics. The interplay between
atomic arrangement and electronic structure is studied in Ref. \cite{Ro04}.
Briefly, donor organic molecules such as BEDT-TTF
(bis-ethylenedithio-tetrathiafulvalene, further abbreviated below as ET),
BEDO-TTF (bis-ethylenedioxy-tetrathiafulvalene) or BEDT-TSF
(bis-ethylenedithio-tetraselenofulvalene) build up conducting planes with
various packing types, so called $\alpha$, $\beta$, $\kappa$, $etc.$ as reported
in Ref. \cite{Sh04}. These planes are separated from each other by insulating,
generally inorganic, acceptor planes. Each donor molecule bears a positive
charge controlled by the anion acceptor charge and the stoichiometry. In numerous
cases, which are of interest in what follows, the unit cell involves 4 donor
molecules with a +1/2 charge yielding two holes per unit cell. FS of these
compounds, which bears similarity with the 3D alkaline-earth metals, originates
from one hole-type orbit (labeled $\bigodot$ in the following) with area equal
to the first Brillouin zone (FBZ) area, that can be approximated as an ellipse
in most cases. In the extended zone scheme, $\bigodot$ orbits overlap and gap
opening is observed at the crossings due to lifting of degeneracy. Depending on
$e.g.$ the strength of $\pi$-$\pi$ interactions between donor molecules, overlap
occurs along either one or two directions, yielding networks of orbits liable to
be connected to each other by magnetic breakdown (MB) in high enough magnetic
field. In the first case, we deal with a linear chain of coupled 2D hole orbits.
Such a network, an example of which is given in Fig.~\ref{Fig:SF_networks}(a), is
an experimental realization of the famous model proposed by Pippard in the early
sixties to study MB \cite{Pi62,Sh84}. In the second case a 2D network of
compensated orbits is observed. This latter network involves two hole-type
orbits labeled $a$ and $b-a$ in Fig.~\ref{Fig:SF_networks}(f) and one
electron-type orbit labeled $b$. Intermediate case is depicted in
Fig.~\ref{Fig:SF_networks}(c) where $\bigodot$ orbits come close together in one
direction. In such a case, a large gap opens around the FBZ boundary and the
network is composed of one hole-type and one electron-type orbit (labeled $a$)
with the same cross section. Analogous type of network is also observed in
(BEDO-TTF)$_2$ReO$_4 \cdot$H$_2$O \cite{Ro04,Kh98} and in compounds with less
trivial FS genesis due to $e.g.$ FBZ folding \cite{Vi08,Ve94} as reported in
Fig.~\ref{Fig:SF_networks}(d). It can also be observed in the Bechgaard salt
(TMTSF)$_2$NO$_3$ (where TMTSF stands for tetramethyl-tetraselena-fulvalene).
This q-1D metal at room temperature becomes q-2D at temperatures below the anion
ordering (T$_{\mathrm {AO}}$ = 41 K). In-between T$_{\mathrm {AO}}$  and the
spin density wave condensation temperature (T$_{\mathrm {SDW}}$ = 9.4 K), its
FS achieves the linear chain of compensated orbits displayed in
Fig.~\ref{Fig:SF_networks}(e) \cite{Kan95}.

Nevertheless, even not to mention structural phase transitions that can strongly affect the electronic structure, experimental data can hardly be reconciled with the calculations of Fig.~\ref{Fig:SF_networks} in few cases. This is mainly due to the extreme sensitivity of organic metals to tiny structural details \cite{Ro04} that can be modified by
external parameters such as  temperature or moderate applied pressure. In that
respect, high magnetic field-induced quantum oscillations are powerful tools for
the study of such FS's. Otherwise, in the numerous cases where the calculations reported in
Fig.~\ref{Fig:SF_networks} holds, MB yields oscillatory features, not predicted
by the semiclassical models, the quantitative interpretation of which is still
in progress.

In Section~\ref{sec:puzzling}, we report on few examples of puzzling features of
the oscillation spectra observed in organic metals with predicted FS such as those
reported in Figs.~\ref{Fig:SF_networks}(c) and (f). The
linear chain of coupled orbits, such as that displayed in Fig.~\ref{Fig:SF_networks}(a), and  compensated electron-hole networks, such as reported in Fig.~\ref{Fig:SF_networks}(d) and (e), are
considered in Sections~\ref{sec:linear_chain} and~\ref{sec:compensated}, respectively, from the viewpoint of the quantum oscillations physics in connection
with MB.

\section{\label{sec:puzzling}Puzzling oscillations spectra}

Shubnikov-de Haas (SdH) effect, $i.e.$ conductivity oscillations, and de
Haas-van Alphen (dHvA) effect, $i.e.$ magnetization oscillations, of multiband
organic metals are generally studied in the framework of the Lifshits-Kosevich model~\cite{LK54,LK55}.
For a single quadratic band $\eta$ of a q-2D system, Landau levels are given by

\begin{equation}
\label{Eq:LL}
E_{\eta,n}=\hbar\omega_{\eta}(n+\frac{1}{2})+\tau_{\perp}\cos(k_z a_{\perp})
\end{equation}

where $n$ is the Landau level index, $\omega_{\eta}=e B\cos\theta /m_{\eta}$ is the cyclotron frequency
($e=|e|$ being the electron charge), $m_{\eta}$ is the
effective mass, $\tau_{\perp}$ is the interlayer energy transfer integral, $a_{\perp}$ is the distance between conducting planes and $\theta$ is
the angle between the direction of the magnetic field $B$ and the normal to the conducting layers. The oscillatory part of the Kubo conductivity, in the approximation of independent collisions, and magnetization
are then given by the semiclassical formulae

\begin{equation}
\label{Eq:SdH}\frac{\sigma-\sigma_0}{2\sigma_0}=\frac{\Delta\sigma}{2\sigma_0}=
\sum_{p \geq
1}\frac{(-1)^p}{p}J_1\Big ( \frac{2\pi p\tau_{\perp}}{\hbar\omega_{\eta}}\Big
)
\Big (
\frac{\hbar\omega_{\eta}}{\pi \tau_{\perp}}+\frac{2\pi p k_BT_D}{ \tau_{\perp}}
\Big )
R_{\eta,p}\cos(2\pi p \frac{F_{\eta}}{B\cos\theta})
\end{equation}

and

\begin{equation}
\label{Eq:dHvA}m_{osc}=-\sum_{p \geq 1}
\frac{(-1)^pF_{\eta}}{\pi p m_{\eta}}
J_0\Big ( \frac{2\pi p\tau_{\perp}}{\hbar\omega_{\eta}}\Big )R_{\eta,p}
\sin(2\pi p\frac{F_{\eta}}{B\cos\theta}),
\end{equation}

respectively, for SdH and dHvA oscillations, where
$\sigma_0=e^2\tau_{\perp}^2a_{\perp}m_{\eta}/(8\pi^2\hbar^3k_BT_D)$ is the
zero-field conductivity (a detailed calculation is given $e. g.$ in Refs.~\cite{PLv10,TH08,Gr03}),
$m_{osc}$ is the oscillatory part of the magnetization~\cite{PLv09}. $T_D$ is the Dingle temperature ($T_{D}$ =
$\hbar$/(2$\pi k_B\tau)$, where $\tau$ is the relaxation time).
The factor $(\hbar\omega_{\eta}/\pi \tau_{\perp}+2\pi p k_BT_D/\tau_{\perp})$ appearing in Eq.~\ref{Eq:SdH} is a contribution from the imaginary part of the retarded Green function $|\Im
G_{\eta,n}(E)|^2$ coming from the Kubo formula, where
$G_{\eta,n}(E)=(E-E_{\eta,n}+i\Delta_{\eta,n}(E))^{-1}$. The squared term, when
expanded, has one pole of first order, giving a contribution in $1/
|\Delta_{\eta,n}(E)|$, and a pole of second order, giving the second
contribution equal to $2\pi p /(\hbar\omega_{\eta})$. In the usual Born
approximation the imaginary part $|\Delta_{\eta,n}(E)|$ is replaced by its
averaged value $<|\Delta_{\eta,n}(E)|>=\pi k_BT_D$ which defines the Dingle
temperature.
Moreover, the oscillating part of the self-energy, which can be
included as corrections of $\pi k_BT_D$, is neglected. In the three-dimensional limit
where $\tau_{perp}\gg \hbar \omega_{\eta}$, these
oscillations may have significant contribution to the SdH effect. In the clean
limit, $\hbar \omega_{\eta}\gg k_B T_D$ they can be discarded. The oscillation
frequency $F_{\eta}=\mu m_{\eta}/(\hbar e)$ of
the classical orbit $\eta$, where $\mu$ is the chemical potential, is
proportional to the cross section area $A_{\eta}$ of the orbit $F_{\eta}$ =
$\hbar A_{\eta}$/(2$\pi e$), which is physically the quantum flux $h/e$ through $A_{\eta}$ divided by $4\pi^2$. The damping factor $R_{\eta,p}$ can be factorized as

\begin{equation}
\label{Eq:damping}
R_{\eta,p}(T,B) =
R^{T}_{\eta,p}R^{D}_{\eta,p}R^{MB}_{\eta,p}R^{s}_{\eta,p}
\end{equation}

where the thermal, Dingle, MB and spin damping factors are given by
the expressions:

\begin{eqnarray}
\label{Eq:RT}R^{T}_{\eta,p} &=&
\frac{u_0Tpm_{\eta}}{B\cos\theta\sinh\Big [
u_0Tpm_{\eta}/(B\cos\theta)\Big ]},
\\
\label{Eq:RD}R^{D}_{\eta,p}&=& \exp\Big
[-u_0T_Dpm_{\eta}/(B\cos\theta)\Big], \\
\label{Eq:RMB}R^{MB}_{\eta,p} &=& (ip_0)^{n^t_{\eta}}(q_0)^{n^r_{\eta}},\\
\label{Eq:Rs}R^{s}_{\eta,p} &=& \cos(\pi p\frac{g^*m_{\eta}}{2\cos\theta}),
\end{eqnarray}

respectively \cite{Sh84}, where $u_0$ = 2$\pi^2m_ek_B/(e\hbar)$ = 14.694 T/K. In the following, effective masses $m_{\eta}$
are expressed in units of the electron mass $m_e$. Integers $n^t_{\eta}$ and
$n^r_{\eta}$ are the number of
MB tunnelings and reflections, respectively, encountered by the quasiparticle
along its closed trajectory. $g^*$ is the effective
Land\'{e} factor. The MB tunneling and reflection probabilities are given
with a good approximation by $p_0$ = $\exp(-B_0/2B)$ and $q_0^2$ = 1 -
$p_0^2$, respectively, where $B_0$ is the MB field. The exact semiclassical expression for the tunneling
parameters can be evaluated by considering more generally the Riemann surface
of the band structure~\cite{Fo98b} which is constructed from a polynomial in the
complex plane $\mathbb{C}^2$. The singularities of this polynomial are essential
in determining the genus of the Riemann surface and its homotopy group. Each
element in the group is associated to a unique complex quantity $S_{\eta}$ (which is a semiclassical action on this surface) whose real part corresponds to $F_{\eta}$ and imaginary
part to the MB field $B_0$, respectively. Even
though direct application of this method is difficult in multiband systems
(see few examples in Ref.~\cite{Fo98b}), it gives a nice framework and clear picture
of the tunneling process.

Furthermore, FS warping can lead to beating features in the oscillatory
spectra due to the interlayer transfer integral $\tau_{\perp}$ entering the Landau spectrum (see
Eq.~\ref{Eq:LL}). Warping is accounted for by the warping factors involving $\tau_{\perp}$ in Eqs.~\ref{Eq:SdH} and~\ref{Eq:dHvA}. However, detailed analysis of its contribution to SdH oscillations
have demonstrated that $\tau_{\perp}$ values of a few tenth of a meV, such as it is
observed in $\beta$-phase compounds \cite{Co89}, yield oscillatory spectra that
cannot be explained by the simple addition of a reduction factor
\cite{Gr03,Ka02,Ka04,Ka05}. Nevertheless, $\tau_{\perp}$ is by one order of magnitude
smaller in numerous q-2D organic metals, leading to warping factor values
close to 1, as far as dHvA oscillations are concerned, and/or slowly varying with the magnetic field.\footnote{For
$\tau_{\perp}$ = 0.04 meV and $m_{\eta}$ = 3.3, which hold for the $\alpha$
orbit of $\kappa$-(ET)$_2$Cu(NCS)$_2$ \cite{Si02}, $J_0(2\pi \tau_{\perp}/\hbar\omega_{\eta})$ varies from 0.88 to 0.98 as the magnetic field increases from 30 to 60 T.}

Finally, it should be mentioned that SdH oscillations are generally measured
through interlayer magnetoresistance (R$_{zz}$) oscillations. In the case where
oscillations amplitude is small compared to the background resistance R$_{bg}$,
it can be assumed that -$\Delta \sigma$/$\sigma_0 \simeq \Delta$R$_{zz}$/R$_{bg}$.
 Besides, contrary to magnetization which, as a thermodynamic parameter, is only
sensitive to the density of states, magnetoresistance oscillations may originate
from quantum interference phenomena (QI) \cite{St71,St74} as early reported for $\kappa$-(BEDT-TTF)$_2$Cu(NCS)$_2$ \cite{Ca94,Ka96,Ha96b}. In such a case,
Eqs.~\ref{Eq:SdH} and~\ref{Eq:RT} to~\ref{Eq:Rs} still hold except that the
effective mass entering Eqs.~\ref{Eq:RT} and~\ref{Eq:RD} is the difference and
sum, respectively, of the two partial effective masses of each of the
interferometer arms. This can lead to small and even zero-effective masses for
symmetric interferometer. In such a case, oscillations are observed up to high
temperature, in a range where the oscillations amplitude damping, still accounted for by
$R^{D}_{\eta,p}$, is due to temperature-dependent relaxation time governed by
inelastic collisions.

\begin{figure}                                                    
\centering
\includegraphics[width=0.9\columnwidth,clip,angle=0]{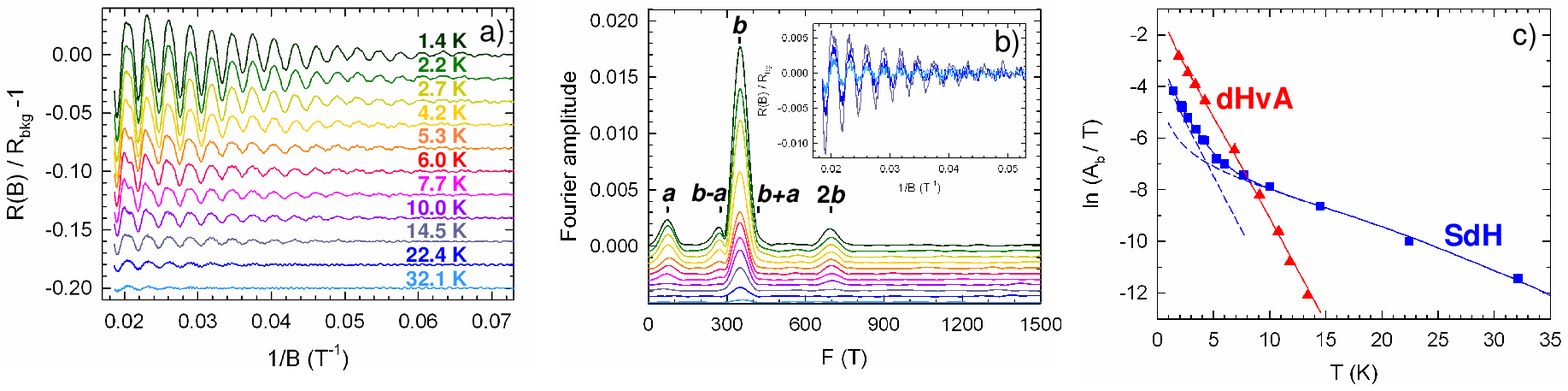}
\caption{\label{Fig:oxalates} (Colour on-line) (a) Oscillatory part of the magnetoresistance data at ambient pressure and (b) corresponding Fourier spectra of $\beta$''-(ET)$_4$(H$_3$O)[Fe(C$_2$O$_4$)$_3$]$\cdot$C$_6$H$_4$Cl$_2$. As displayed in the inset of (b), SdH oscillations are still observed at 32 K. Curves in main panels of (a) and (b), are shifted down from each other by a constant amount. (c) Temperature dependence of the $b$
oscillations amplitude at 30 T and 34.3 T ($i.e.$ 30 T/cos(29$^{\circ}$)), respectively, for SdH and dHvA data collected at ambient pressure. Solid squares are SdH data deduced from (a); solid triangles are dHvA data for $\theta$ = 29$^{\circ}$. Solid lines are the best fits of the LK model assuming either only one closed orbit contributes to the amplitude (dHvA data) or the coexistence of a closed orbit and a quantum interferometer with a zero-effective mass and a temperature-dependent scattering rate (SdH data). Each of these two contributions are displayed as dashed lines. From Refs.~\cite{Vi09,La11}.}
\end{figure}

In the following, we will consider a family of organic conductors illustrating the sensitivity of the electronic structure, hence of the quantum oscillations spectra, of organic metals to small changes of the atomic structure induced by either chemical substitutions or moderate applied pressure.

Charge transfer salts with the generic formula $\beta$''-(ET)$_4$(A)[M(C$_2$O$_4$)$_3$]$\cdot$Solv (where A is a monovalent cation, M is a trivalent cation and Solv is a solvent) have raised great interest for many years \cite{Cor04}. These salts which share the same $\beta$'' packing of the ET molecules can be either orthorhombic, in which case they are insulating, or monoclinic q-2D metals. Among these latter salts, denoted as (A, M, Solv) hereafter, many different ground-states, including normal metal, charge density wave, superconductivity, and temperature-dependent behaviours can be observed. The FS of (NH$_4$, Fe, C$_3$H$_7$NO) is displayed in Fig.~\ref{Fig:SF_networks}(c) where the area of the compensated orbits amounts to
8.8~$\%$ of the FBZ area \cite{Pr03}. Analogous FS is also predicted for (H$_3$O, Fe, C$_6$H$_5$CN) \cite{Ku95}. In qualitative agreement with these calculations, only one Fourier component with frequency F = 230 T, corresponding to 6 $\%$ of the FBZ area, is observed in SdH spectra of (H$_3$O, Ga, C$_6$H$_5$NO$_2$) \cite{Ba04}. However, the FS of other compounds of this family can be more complicated since four frequencies corresponding to orbits area in the range 1.1 to 8.5~$\%$ of the FBZ area are observed for (H$_3$O, M, C$_5$H$_5$N) where M = Cr, Ga, Fe \cite{Co04}. In this latter case, a density wave ground state, responsible for the observed strongly non-monotonous temperature dependence of the resistance, has been invoked to account for this discrepancy. In contrast, only two frequencies are observed for (H$_3$O, M, C$_6$H$_5$NO$_2$) where M = Cr, Ga \cite{Ba05}. Besides, moderate applied pressure have a drastic effect on the SdH oscillations spectra of (NH$_4$, Cr, C$_3$H$_7$NO). Namely, whereas up to 6 Fourier components corresponding to orbit area in the range 0.1~$\%$ to 7~$\%$ of the FBZ area are observed at ambient pressure, the spectrum simplifies under moderate pressure since only 3 components (F$_1$ = 68 $\pm$ 2 T, F$_2$ = 238 $\pm$ 4 T, F$_3$ = 313 $\pm$ 7 T) corresponding to orbits area in the range 1.7 $\%$ to 7.8 $\%$ of the FBZ area are observed at 1 GPa \cite{Vi06}. These scattered results demonstrate the sensitivity of the electronic structure of organic metals, hampering any interpretation in the framework of the band structure calculations. The last result could nevertheless be interpreted on the basis of the data in Fig.~\ref{Fig:SF_networks}. Indeed, as pointed out in Ref.~\cite{Pr03}, the $\bigodot$ orbits of Fig.~\ref{Fig:SF_networks}(c)  may also intersect along the small axis, leading to one additional orbit as reported in Fig.~\ref{Fig:SF_networks}(f). In such a case, 3 Fourier components linked by a linear combination settled by the orbits compensation, F$_b$ = F$_a$ + F$_{b-a}$, should be observed. This picture holds, not only for the above compound under pressure for which F$_3$ = F$_1$ + F$_{2}$ within the error bars, but also for (NH$_4$, Fe, C$_3$H$_7$NO) in the applied pressure range from ambient pressure to 1 GPa \cite{Au04,Au06} and for (H$_3$O, Fe, C$_6$H$_4$Cl$_2$) \cite{Vi09,La11,Vi10}.

Further information can be obtained from the field and temperature dependence of the oscillatory spectra (see Eqs.~\ref{Eq:SdH} to~\ref{Eq:Rs}). SdH and dHvA oscillations of the latter compound have been considered as displayed in Fig.~\ref{Fig:oxalates}. Whereas  $a$ and $b-a$ SdH oscillations follow the LK behaviour, a kink is observed in the mass plot of the $b$ oscillations at $\sim$ 7 K. As a result, despite a rather large Dingle temperature (T$_D$ = 4 $\pm$ 1 K), SdH oscillations are still observed at 32 K which, to our knowledge, constitutes a world record for an organic metal. In contrast, the dHvA oscillations are in agreement with the LK formula in all the explored temperature range as displayed in Fig.~\ref{Fig:oxalates}(c). Keeping in mind that dHvA oscillations are only sensitive to the density of states, it has been shown that both the field and temperature dependence of the $b$ oscillations are consistently accounted for by the coexistence of a closed $b$ orbit and a quantum interferometer with the same cross section \cite{Vi09,La11,Vi10}. Unfortunately, existence of a QI path is not consistent with the FS of  Fig.~\ref{Fig:SF_networks}(f) and this amazing result therefore remains unexplained.

Otherwise, the frequency F$_{b+a}$ is observed both for  (NH$_4$, Fe, C$_3$H$_7$NO)
\cite{Au04,Au06} and (H$_3$O, Fe, C$_6$H$_4$Cl$_2$) under pressure
\cite{La11,Vi10}. Taking into account the opposite sign of the $a$ and $b$
orbits, the $b+a$ component cannot correspond to a MB orbit \cite{Fa66} in the
framework of the FS of Fig.~\ref{Fig:SF_networks}(f). In addition, its effective
mass is lower than both m$_{a}$, m$_{b-a}$ and m$_{b}$ which is at odds
with the Falicov-Stachowiak model as well. Therefore, this component has not a
semiclassical origin and can be attributed to the frequency combination
phenomenon considered in the next sections.

\section{\label{sec:linear_chain}The linear chain of coupled orbits}

\begin{figure}
\subfigure[ ]
{\includegraphics[scale=0.45]{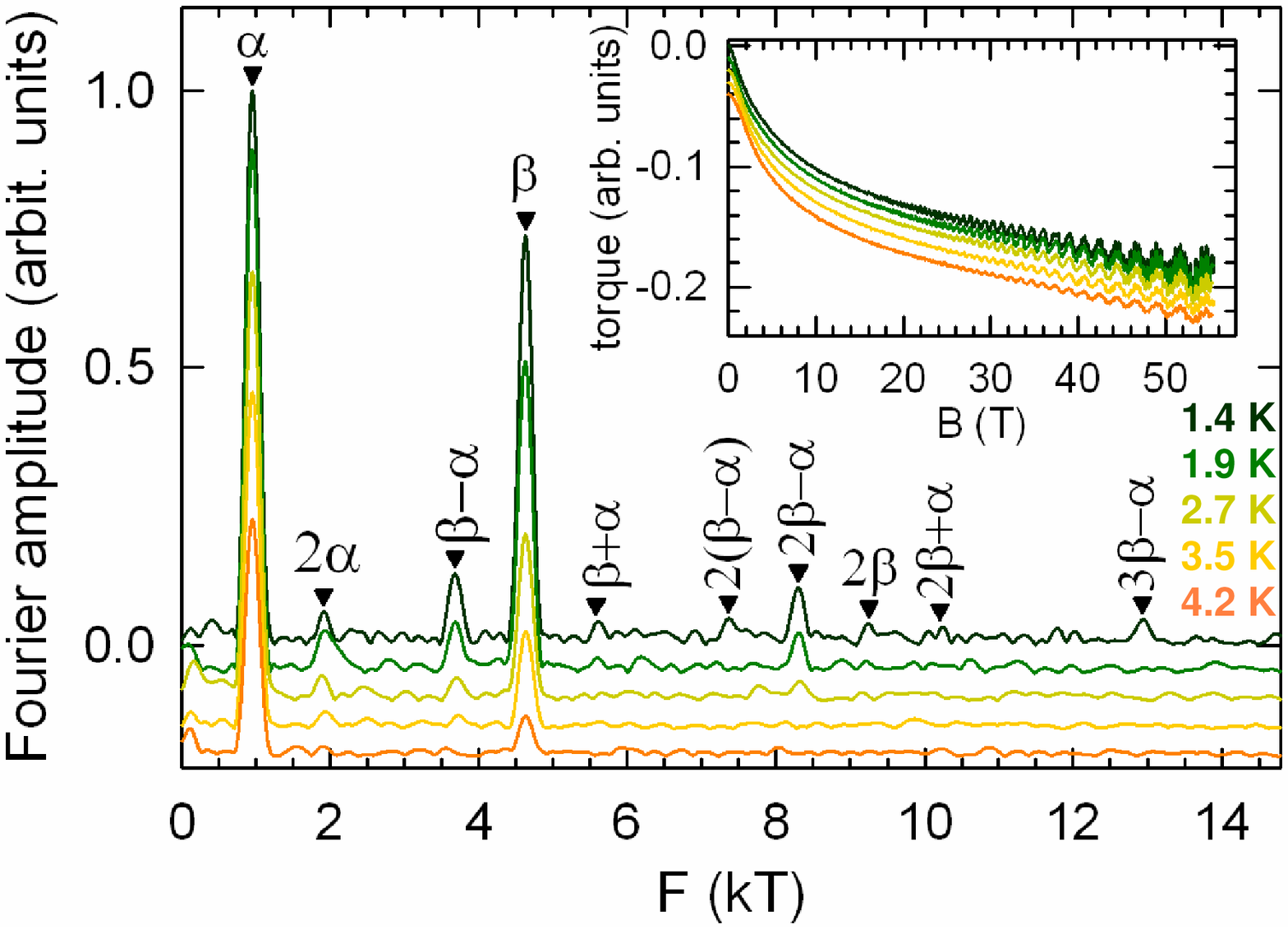}
\label{Fig:mm_TF_torque}}
\subfigure[ ]
{\includegraphics[scale=0.45]{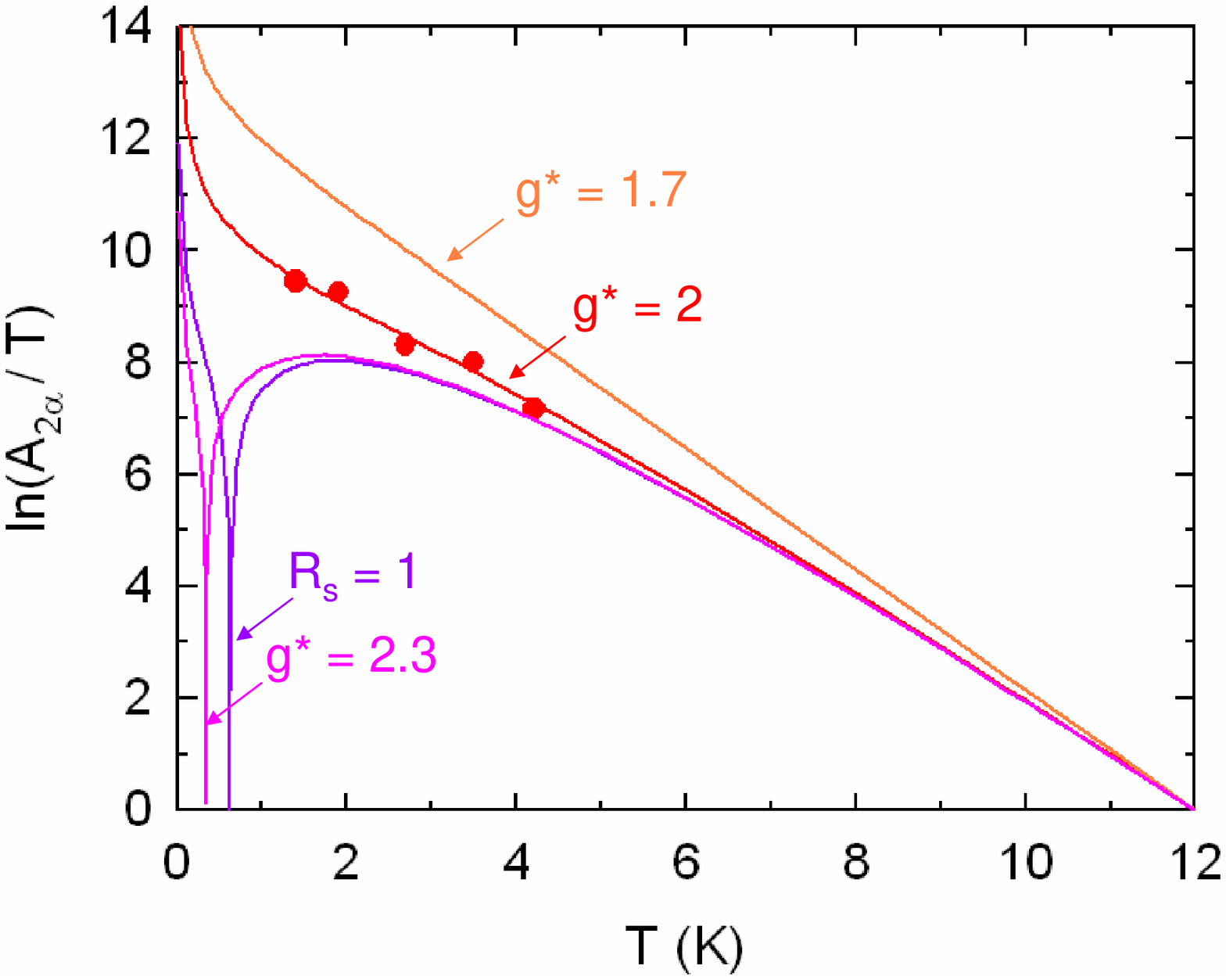}
\label{Fig:mass_plot_2a}}
\caption{ (Color on-line) (a) Fourier analysis in the field range 35-55.3 T of the magnetic torque data, relevant to $\theta$-(ET)$_4$CoBr$_4$(C$_6$H$_4$Cl$_2$), reported in the insert. Solid triangles are marks calculated with F$_{\alpha}(\theta = 0)$ = 0.944 kT and F$_{\beta}(\theta = 0)$ = 4.60 kT. Data have been shifted down from each
other by a constant amount.
(b) Temperature dependence (mass plot) of the 2$\alpha$ amplitude. Symbols are
data deduced from torque measurements of Fig.~\ref{Fig:mm_TF_torque}
at 48.3 T. Solid lines are deduced from Eq.~\ref{Eq:A2alpha} with $m_\alpha$ =
1.81, $m_\beta$ = 3.52,
$T_D$ = 0.79 K and $B_0$ = 35 T. Spin damping factor (see Eq.~\ref{Eq:Rs}) is
either neglected ($R^s_p$ = 1, p = 1 to 3) or
accounted for by various $g^*$ values. All the data are normalized to their
values at 12 K. }
\end{figure}

MB phenomenon has been intensively studied in 3D alkaline-earth metals over the
sixties and seventies (see \cite{Sh84} and references therein). In order to
compute MB, Pippard introduced the linear chain of coupled orbits in the early
sixties \cite{Pi62}. The energy spectrum of this model FS, $i. e.$ the density of states in magnetic field in which all the allowed classical orbits contribute, can be computed analytically. In contrast to the semiclassical picture, a set of Landau bands broadened by coherent MB is obtained instead of discrete Landau levels, due to the traveling of quasiparticles on the non-quantized q-1D sheets of the FS. However, the Falicov-Stachowiak semiclassical model \cite{Fa66}, based on the LK model and involving discrete Landau levels, yielded analytic tool that successfully accounted for the field
and temperature dependence of the oscillations spectra of these multiband metals
\cite{Sh84}. This success outshined the Pippard result until the discovery of
frequency combinations, 'forbidden' in the framework of the Falicov-Stachowiak
model, in magnesium \cite{Ed82} and, later on, the experimental realization of
the FS proposed by Pippard in the starring compound
$\kappa$-(ET)$_2$Cu(NCS)$_2$ \cite{Os88}. Many other organic metals share this
FS topology, an example of which is given in Fig.~\ref{Fig:SF_networks}(a). The
main feature of the dHvA spectra of these compounds is the presence of
the forbidden Fourier component $\beta - \alpha$ \cite{Me95,Uj97,St99}.
Furthermore, depending on the field and temperature range explored, effective
masses relevant to frequency combinations linked to MB orbits can be at odds
with the predictions of the Falicov-Stachowiak model, both for dHvA
\cite{Uj97,St99} and SdH \cite{Ha96b,Vi07} oscillations, even taking into account
QI in the latter case. Numerical and analytical analysis of this spectrum yield, through numerical resolution, a non-LK behaviour of the oscillations amplitude~\cite{Ha96,Sa97,Fo98a}. Besides, the forbidden Fourier components
observed in these compounds result not only from the formation of Landau bands
but are also due to the oscillation of the chemical potential
enabled by the 2D character of the FS
\cite{Ha96,Fo98a,Al96,Al01,Ch02,Ki02,Fo05}.

Comprehensive analytical calculations of the oscillations amplitude, taking into account both coherent MB and oscillation of the chemical potential, have been reported in Refs.~\cite{Gv02,Gv04} and
implemented in dHvA data of $\kappa$-(ET)$_2$Cu(NCS)$_2$. However, easy to handle analytic tools necessary
to quantitatively account for the experimental data were still lacking. Indeed,
as pointed out in Ref.~\cite{Gv04}, the equations are complex and the data could be
analyzed only numerically.
In Ref.~\cite{Au12}, a model based on the field-induced chemical
potential oscillations, yielding analytical non-LK
expressions for the Fourier components amplitude is proposed. This model accounts for the temperature- and field-dependent quantum oscillations spectra of the strongly
two-dimensional charge transfer salt
$\theta$-(ET)$_4$CoBr$_4$(C$_6$H$_4$Cl$_2$) with an excellent agreement. Remarkably, the unit cell of this compound involves two different donor layers~\cite{Sh11}. One of them is insulating whereas the FS of the other, displayed in Fig.~\ref{Fig:SF_networks}(a), is an illustration of the Pippard's model. Fourier spectra of the
dHvA oscillations displayed in Fig.~\ref{Fig:mm_TF_torque} reveal the
existence of both classical orbits $\alpha$, $\beta$, $2\beta-\alpha$ and
forbidden frequencies such as $\beta-\alpha$ and its harmonics.

Considering a two band system with effective masses $m_{0(1)}$ and band extrema $\Delta_{0(1)}$, as displayed in
Fig.~\ref{Fig:SF_networks}(b), the grand
potential $\Omega$ of a 2D slab with area $\mathcal{A}$ is given by
a series of harmonics $p$ of classical frequencies $F_\eta$

\begin{eqnarray}\nonumber
\frac{\Phi_0}{\mathcal{A}}\frac{u_0}{k_B}\Omega(T,\mu)&=&
\tOmega(T,\mu)=-\frac{
m_0}{2}(\mu-\Delta_0)^2
-\frac{m_1}{2}(\mu-\Delta_1)^2
\\ \label{Eq:Omega}
&+&\frac{(B\cos\theta)^2}{2}\sum_{p\geq1}\sum_{\eta}\frac{C_{\eta}}{\pi^2p^2m_{
\eta } }
R_{\eta,p}(T,B)\cos(2\pi p\frac{F_{\eta}}{B\cos\theta}+p\varphi_{\eta})
\end{eqnarray}

in the low temperature limit and within the semiclassical approximation. The chemical potential $\mu$ and energies $\Delta_{0(1)}$ are expressed in Tesla. The universal coefficient $\Phi_0 u_0/k_B$ (where $\Phi_0$ = $h/e$ is the quantum flux), in unit of which the physical quantities are expressed for simplification, is equal to $4\pi^3m_e/e^2\simeq 4.4\times 10^9$ T$^2$m$^2$/J.
$\eta$ stands for the fundamental closed orbits that are not
harmonics but can be composed
of different parts of the FS thanks to MB. In particular, two fundamental
orbits with frequencies $F_{\alpha}$ = $m_1$($\mu$ - $\Delta_{1}$)
(i.e.  $m_{\alpha}$ = $m_1$) for the smallest and $F_{\beta}$ = $m_0$($\mu$ -
$\Delta_{0}$) + $m_1$($\mu$ - $\Delta_{1}$), which
can be written as $F_{\beta}$ = $m_{\beta}$($\mu$ - $\Delta_{\beta}$), for the
largest are defined (see Fig.~\ref{Fig:SF_networks}(a)). Within this framework,
$m_0$ + $m_1$ is identified with the effective mass $m_{\beta}$ of the MB orbit
$\beta$. Each orbit $\eta$ has an effective mass $m_{\eta}$ and yields the frequency $F_{\eta}$ =
$m_{\eta}$($\mu$ - $\Delta_{\eta}$) where $m_{\eta}$ and $\Delta_{\eta}$ are
combinations of the fundamental parameters $m_{0,1}$ and $\Delta_{0,1}$. $\varphi_{\eta}$ is the phase or Maslov index
determined by the number of
turning points on the trajectory $\eta$. At each turning point is associated a
value $\pi/2$, yielding $\varphi_{\alpha}=\varphi_{\beta}=\pi$ for  orbits $\alpha$ and $\beta$ of  Fig.~\ref{Fig:SF_networks}(a), which is the value for a parabolic band model near the Fermi energy.
For more complicated orbits, $\varphi_{\eta}$ is a multiple of $\pi$. For
example the phase of the fundamental orbit $\alpha + \beta$ is $\varphi_{\alpha+\beta}=2\pi$.
$C_{\eta}$ is the symmetry factor of the orbit $\eta$, which counts the
number of non-equivalent possibilities for an orbit to be drawn on the FS. For
example, $C_{\alpha}=C_{\beta}=C_{2\beta-\alpha}=1$ and $C_{\beta+\alpha}=2$.
For a FS composed of a single orbit, $C_{\eta}=1$.
We notice that frequencies $F_{\eta}$ depend explicitly on the chemical
potential $\mu$. Indeed, a frequency physically represents a filling factor proportional to the orbit area in the Brillouin zone or, equivalently, the occupation number of the quasiparticles.
For example, $F_{\beta}$ corresponds to the total area of the FBZ. Eq.~\ref{Eq:Omega} can be easily extended to more complex multiband systems. It is derived from the usual semi-classical technique using the Poisson formula
applied in the case where $\hbar\omega_c$ is small compared to the chemical potential $\mu$~\cite{PLv09}. The oscillating term proportional to $B^2$ in Eq.~\ref{Eq:Omega} contains all the possible contributions of
MB between the bands. In the Grand Canonical Ensemble, the chemical potential is
independent of the magnetic field,
leading to the expression of the magnetization $m=-\partial\tOmega/\partial
(B\cos \theta)$, hence Eq.~\ref{Eq:dHvA}. In the case where the
electron density is fixed, which is common in dHvA experiments, the chemical
potential generally depends on the magnetic field and oscillates. These
oscillations can be damped if the system is connected to a reservoir of uniform
electron density~\cite{Ha96}.
The electron density per surface area $n_e$ is defined by d$\tOmega$/d$\mu$ =
-$n_e$. In zero-field, Eq.~\ref{Eq:Omega} yields $n_e$ = ($m_0$ + $m_1$)$\mu_0$
- $m_0\Delta_0$ - $m_1\Delta_1$ where $\mu_0$ is the zero-field chemical
potential. In the presence of a magnetic field, $\mu$ satisfies instead the
following implicit equation

\begin{eqnarray}
\label{Eq:mu}
\mu=\mu_0-\sum_{p\geq1}\sum_{\eta}\frac{B\cos\theta}{m_{\beta}\pi p}
C_{\eta}R_{\eta,p}(T,B)\sin( \frac{2\pi
pF_{\eta}}{B\cos\theta}+p\varphi_{\eta}).
\end{eqnarray}

The small oscillating part of the magnetization can be computed systematically
by inserting Eq.~\ref{Eq:mu} in the $\mu$-dependent terms of Eq.~\ref{Eq:Omega}, in particular the frequencies, and computing $m=-\partial\tOmega/\partial
B\cos(\theta)$. A numerical resolution of this resulting implicit equation could be done to obtain recursively the field dependence of $\mu$. Nevertheless, a more user-friendly controlled expansion in powers of the damping factors
$R_{\eta,p}$ can be derived systematically at any possible order. Up
to the second order, the following analytical expression is obtained for the
oscillating part of the magnetization:

\begin{eqnarray}
\nonumber
& &\label{Eq:mosc}m_{osc}=-\sum_{\eta}\sum_{p\ge 1}\frac{F_{\eta}C_{\eta}}{\pi p
m_{\eta}}
R_{\eta,p}(T)\sin\left ( \frac{2\pi pF_{\eta}}{B\cos\theta}+p\varphi_{\eta}
\right )
+\sum_{\eta,\eta'}\sum_{p,p'\ge 1}\frac{F_{\eta}C_{\eta}C_{\eta'}}{\pi p'
m_{\beta}}
R_{\eta,p}(T)R_{\eta',p'}(T)\times
\\ 
& &\left [
\sin\left ( 2\pi
\frac{pF_{\eta}+p'F_{\eta'}}{B\cos\theta}+p\varphi_{\eta}+p'\varphi_{\eta'}
\right )-
\sin\left ( 2\pi
\frac{pF_{\eta}-p'F_{\eta'}}{B\cos\theta}+p\varphi_{\eta}-p'\varphi_{\eta'}
\right )
\right ]+\cdots
\end{eqnarray}

where frequencies $F_{\eta}=m_{\eta}(\mu_0-\Delta_{\eta})$ are evaluated at $\mu$ = $\mu_0$. Eq.~\ref{Eq:mosc} can be written as a
sum of periodic functions $m_{osc}=\sum_{i}A_i\sin(2\pi F_i/(B\cos\theta))$
where
the index $i$ stands for either classical orbits
$\eta$ or forbidden orbits such as $\beta$ - $\alpha$. For the most relevant amplitudes, from the experimental data viewpoint, we obtain the following expressions:

\begin{eqnarray}
\label{Eq:Aalpha}A_{\alpha}&=&\frac{F_{\alpha}}{\pi m_{\alpha}}R_{\alpha,1}
+\frac{F_{\alpha}}{2\pi m_{\beta}}R_{\alpha,1}R_{\alpha,2}
+\frac{F_{\alpha}}{6\pi m_{\beta}}R_{\alpha,2}R_{\alpha,3}
+\frac{2F_{\alpha}}{\pi m_{\beta}}R_{\beta,1}R_{\alpha+\beta,1},
\\
\label{Eq:A2alpha}A_{2\alpha}&=&-\frac{F_{\alpha}}{2\pi m_{\alpha}}R_{\alpha,2}+
\frac{F_{\alpha}}{\pi m_{\beta}}R_{\alpha,1}^2-\frac{2F_{\alpha}}{3\pi
m_{\beta}}R_{\alpha,1}R_{\alpha,3},
\\
\label{Eq:Abeta}A_{\beta}&=&\frac{F_{\beta}}{\pi m_{\beta}}R_{\beta,1}
+\frac{F_{\beta}}{2\pi m_{\beta}}R_{\beta,1}R_{\beta,2}
+\frac{F_{\beta}}{6\pi m_{\beta}}R_{\beta,2}R_{\beta,3}
+\frac{2F_{\beta}}{\pi m_{\beta}}R_{\beta,1}R_{\alpha+\beta,1},
\\
\label{Eq:A2beta}A_{2\beta}&=&-\frac{F_{\beta}}{2\pi m_{\beta}}R_{\beta,2}+
\frac{F_{\beta}}{\pi m_{\beta}}R_{\beta,1}^2-\frac{2F_{\beta}}{3\pi
m_{\beta}}R_{\beta,1}R_{\beta,3},
\\
\label{Eq:Abeta-alpha}A_{\beta-\alpha}&=&-\frac{F_{\beta-\alpha}}{\pi
m_{\beta}}R_{\alpha,1}R_{\beta,1}
-\frac{F_{\beta-\alpha}}{\pi m_{\beta}}R_{\alpha,2}R_{\alpha+\beta,1}
-\frac{F_{\beta-\alpha}}{\pi m_{\beta}}R_{\beta,2}R_{\alpha+\beta,1},
\\
\label{Eq:Abeta+alpha}A_{\beta+\alpha}&=&-\frac{2F_{\beta+\alpha}}{\pi
m_{\beta+\alpha}}
R_{\beta+\alpha,1}+\frac{F_{\beta+\alpha}}{\pi
m_{\beta}}R_{\alpha,1}R_{\beta,1},
\\
\label{Eq:A2beta-alpha}A_{2\beta-\alpha}&=&\frac{F_{2\beta-\alpha}}{\pi
m_{2\beta-\alpha}}
R_{2\beta-\alpha,1}+
\frac{F_{2\beta-\alpha}}{2\pi m_{\beta}}
R_{\alpha,1}R_{\beta,2}+
\frac{F_{2\beta-\alpha}}{6\pi m_{\beta}}
R_{\alpha,3}R_{\alpha+\beta,2}.
\end{eqnarray}

These equations differ from the LK model in the sense that a
basic orbit such as $\alpha$, which should involve only one damping factor
$A_{\alpha}=F_{\alpha}R_{\alpha}/(\pi m_{\alpha})$ according to Eq.~\ref{Eq:dHvA}, involves additional terms which are
power combination of different damping factors. However, deviations from the LK
behaviour due
to the high order terms are significant only in the low temperature and
high field ranges. This statement also stands for the classical
orbit 2$\beta$-$\alpha$ since $R_{2\beta-\alpha}$ is significantly higher than
the product $R_{\alpha,1}R_{\beta,2}$ (see Eq.~\ref{Eq:A2beta-alpha}).
Eq.~\ref{Eq:mosc} bring out forbidden frequencies such as
$\beta-\alpha$ (see Eq.~\ref{Eq:Abeta-alpha}). The amplitude of such Fourier component arises from the
combinations of classical orbits $\alpha$ and $\beta$, hence only at the
second order, through the damping factor product $R_{\alpha,1}R_{\beta,1}$.  More generally, the main rule
that can be derived is that products of damping
factors involved at a given order represent algebraic combinations of
frequencies. For example, the orbit $\beta$ in
Eq.~\ref{Eq:Abeta} can be viewed as the combinations $2\beta$ -
$\beta$, or $3\beta$ - $2\beta$, yielding the factors  $R_{\beta,2}R_{\beta,1}$
and $R_{\beta,3}R_{\beta,2}$, respectively, as additional factors entering the
amplitude $A_{\beta}$. As a consequence, forbidden
frequencies arise from the order two since they
cannot be due to any single classical orbit.
Algebraic sums of damping terms in Eqs.~\ref{Eq:A2alpha}
and~\ref{Eq:Abeta+alpha} may possess minus signs which account for $\pi$
dephasings, and may cancel at field and temperature values, depending
strongly on the effective masses, Dingle temperature, MB field, $etc.$
as displayed in Fig.~\ref{Fig:mass_plot_2a} relevant to 2$\alpha$. This point
has already been reported for the second harmonic of the basic orbits both for
compensated \cite{Fo09} and un-compensated \cite{Fo05} metals.

Finally, even taking into account the contribution of QI, stronger deviations from the semiclassical model are reported in the case of SdH spectra of several compounds with this FS topology \cite{Au12,Ha96,Vi07}. These features are not accounted for by the above calculations which therefore requires a specific model.

\section{\label{sec:compensated}Networks of compensated orbits}

\begin{figure} [h]                                                    
\centering
\includegraphics[width=0.4\columnwidth,clip,angle=0]{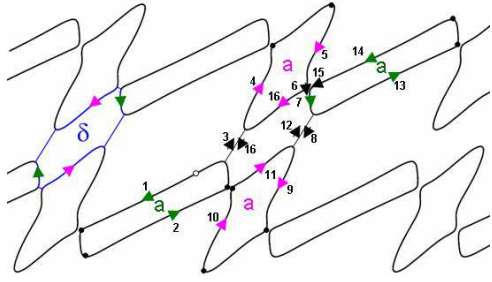}
\caption{\label{Fig:delta} (Colour on-line) Forbidden (left side) and MB-induced (right side) $\delta$ orbit relevant to the Fermi surface of (ET)$_8$[Hg$_4$Cl$_{12}$(C$_6$H$_5$Br)$_2$] \cite{Vi08}. Numbers on the right side indicates the successive steps of the quasiparticles starting from an (arbitrary) origin marked by the open circle. Arrows mark the quasiparticles path. Pink and green arrows correspond to hole and electron parts, respectively, of the Fermi surface. Black arrows and circles mark the magnetic breakdown tunneling junctions and reflections, respectively. Taking into account the opposite sign of the compensated electron and hole basic orbits (labeled $a$), the considered MB orbit yields a frequency corresponding to the area of $\delta$.}
\end{figure}

\begin{figure}                                                    
\centering
\includegraphics[width=0.9\columnwidth,clip,angle=0]{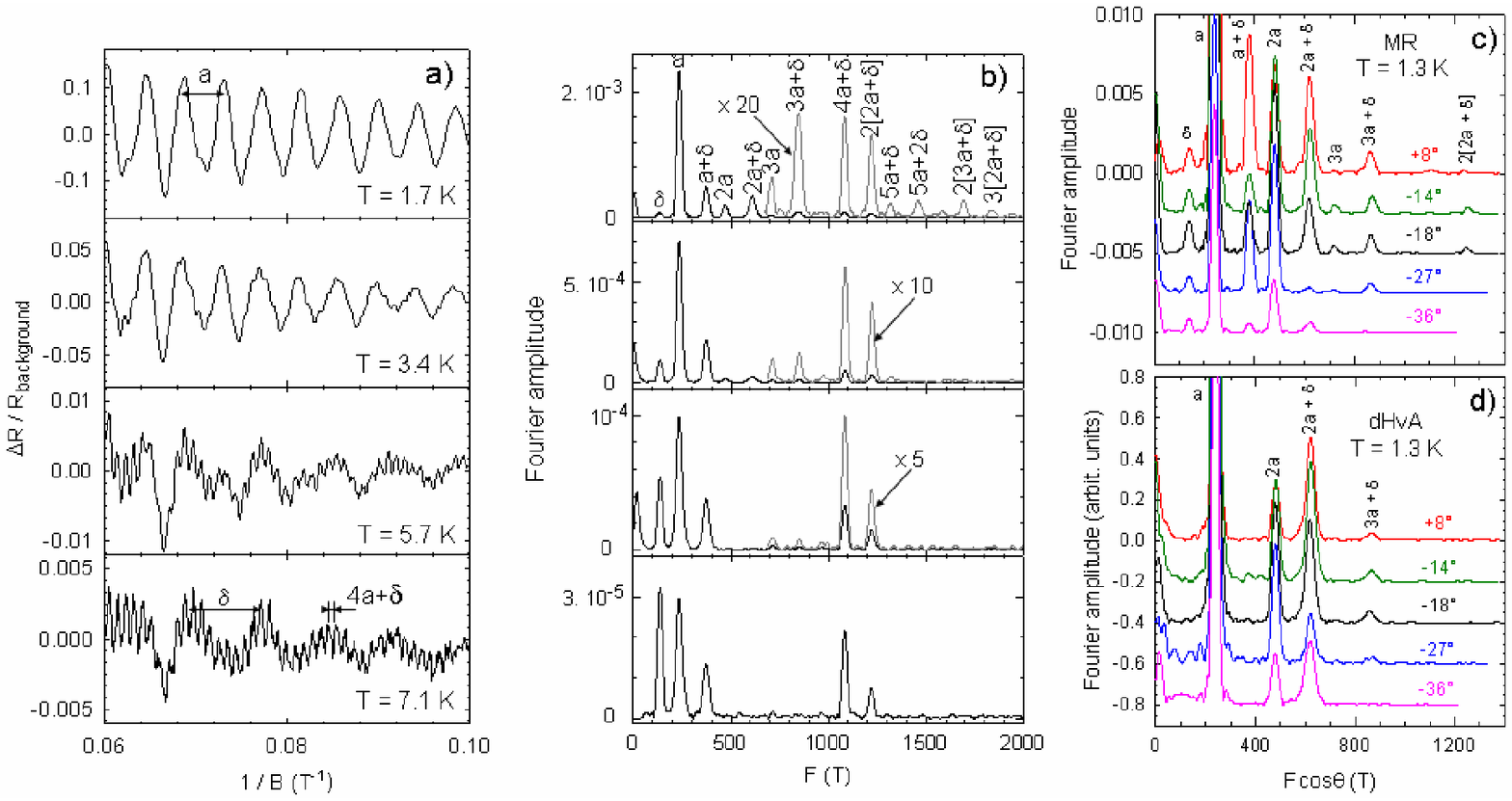}
\caption{\label{Fig:ClBr} (Colour on-line) (a) Oscillatory part of the
magnetoresistance and (b) corresponding Fourier spectra  of
(ET)$_8$[Hg$_4$Cl$_{12}$(C$_6$H$_5$Br)$_2$] at various temperatures \cite{Vi03}.
 Fourier spectra of (c) magnetoresistance and (d) de Haas-van Alphen
oscillations for various directions of the magnetic field \cite{Au05}. Data in
(c) and (d) have been measured simultaneously. Labels $a$ and $\delta$
correspond to closed and forbidden orbits, respectively, depicted in
Fig~\ref{Fig:SF_networks}(d). Contrary to magnetoresistance oscillations spectra
which exhibits forbidden frequency combinations, all the observed dHvA components are
accounted for by the Falicov-Stachowiak semiclassical model.}
\end{figure}

Let us consider now the other class of FS's, widely encountered in organic metals, which are built up with
compensated electron and hole orbits (few examples are reported in
Figs.~\ref{Fig:SF_networks}(c)-(f)). Depending on MB gaps value, either isolated
orbits, 1D or 2D networks are observed. As mentioned in Section
\ref{sec:puzzling}, few hints of forbidden frequency combinations are observed
in SdH spectra of
$\beta$''-(ET)$_4$NH$_4$[Fe(C$_2$O$_4$)$_3$]$\cdot$C$_3$H$_7$NO. Besides, QI
which requires MB gaps crossing as well, have been tentatively inferred to account for
the oscillatory data of
$\beta$''-(ET)$_4$(H$_3$O)[Fe(C$_2$O$_4$)$_3$]$\cdot$C$_6$H$_4$Cl$_2$ reported
in Fig~\ref{Fig:oxalates}. However, to our knowledge, the only compounds
studied from this viewpoint belong to the family
(ET)$_8$[Hg$_4$X$_{12}$(C$_6$H$_5$Y)$_2$] (where X, Y = Cl, Br)
\cite{Vi08,Ly96,Pr02,Vi03,Au05,Au11}. Indeed, due to moderate MB gaps, compounds with X
= Cl achieves 2D networks as reported in Fig.~\ref{Fig:SF_networks}(d). In addition, small scattering rate (T$_D$ = 0.2 $\pm$ 0.2 K) is observed \cite{Vi03}. In line
with the Falicov-Stachowiak model, Fourier components are linear combinations of
the frequencies linked to the compensated closed orbits $a$ and to the
'forbidden' orbits $\delta$ and $\Delta$ \cite{Vi08,Pr02,Vi03,Au05}. This is the
case of $e. g.$ $2a + \delta$ which is observed both in SdH and dHvA spectra.
Regarding magnetoresistance, components linked to QI such as $a + \delta$ and
$2a + \delta + \Delta$ are also observed in the spectra. Strictly speaking, there is no forbidden orbit in a
2D network of compensated orbits. For example, keeping in mind that electron and hole orbits have opposite signs, a MB-induced orbit with the frequency corresponding to $\delta$ can be defined for the FS of Fig.~\ref{Fig:SF_networks}(d) as displayed in Fig.~\ref{Fig:delta}. However, according to the Falicov-Stachowiak
model, its amplitude should be very small. Indeed, this orbit involves 6 MB tunnelings and 10 reflections leading to small $R^{MB}$ values (see Eq.~\ref{Eq:RMB}). Besides, its effective mass $m_{\delta}$ should amount to 4$m_a$
leading to small $R^D$ and $R^T$ values, as well. In violation of these statements, 'forbidden'
orbits such as $\delta$ or $4a + \delta$, with a small effective mass ($m_{\delta} \simeq$ 0.4$m_a$, $m_{4a+\delta} \simeq$ 0.3$m_a$ \cite{Vi03}) and a large amplitude, are
observed. As a result, these two Fourier
components which should not be observed in the framework of the semiclassical
model, are preponderant above few kelvins at moderate fields (see Fig.~\ref{Fig:ClBr}).

\begin{figure}
\subfigure[ ]
{\includegraphics[scale=0.4]{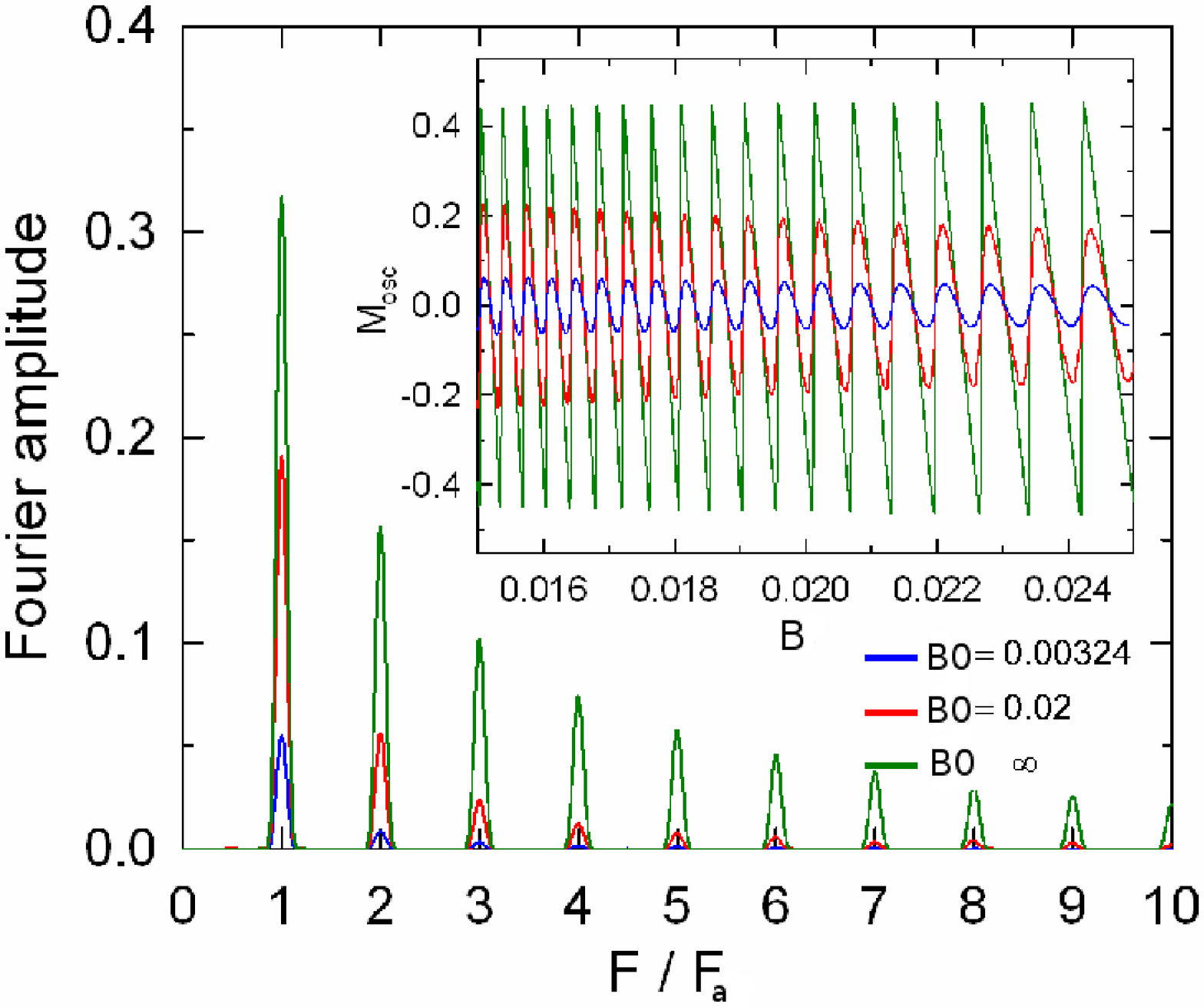}
\label{Fig:TFcomp}}
\subfigure[ ]
{\includegraphics[scale=0.4]{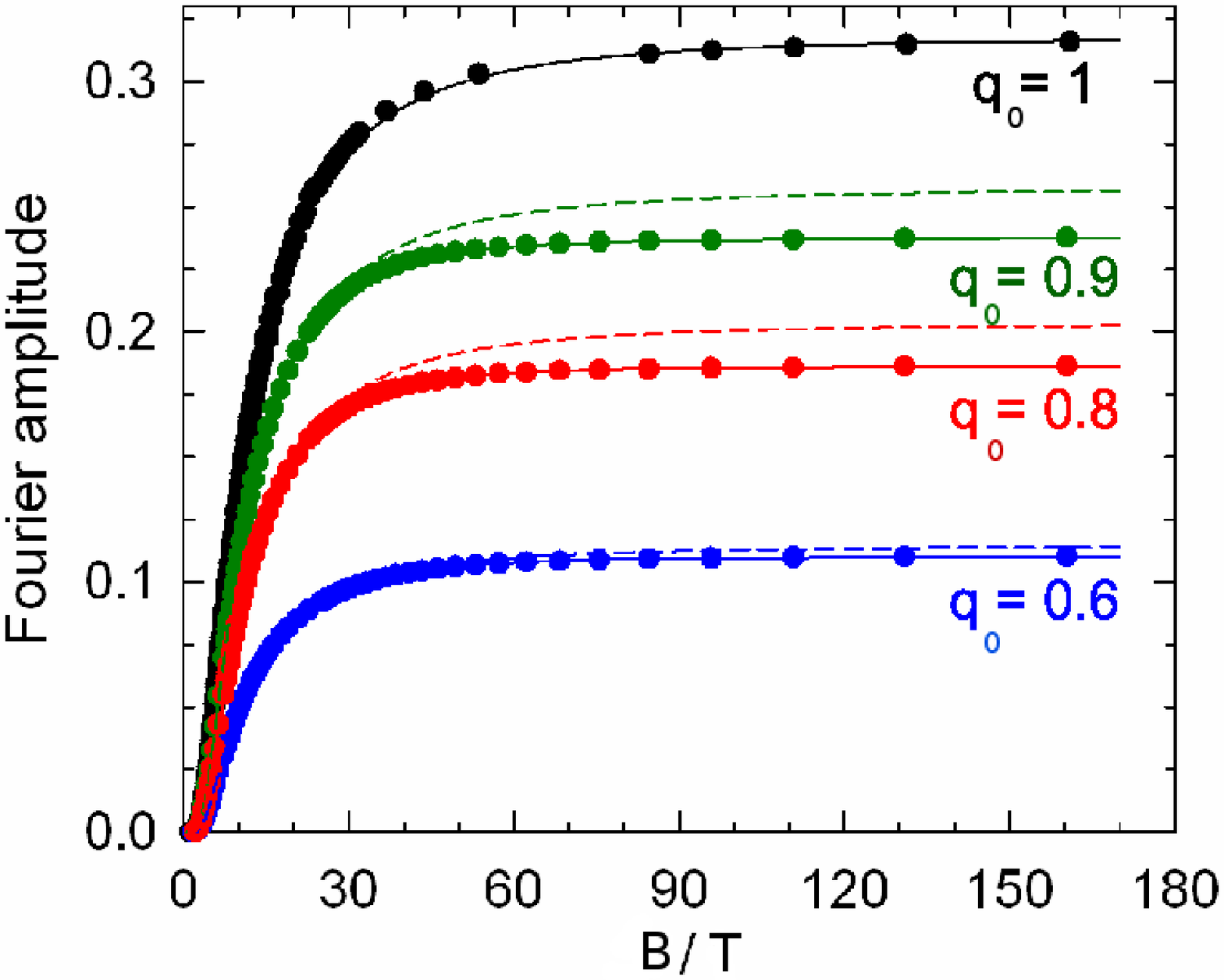}
\label{Fig:h1}}
\caption{ (Color on-line) (a) Fourier spectra of the dHvA oscillations, displayed in the insert, of a linear
chain of compensated orbits, for various values of the magnetic breakdown field
$B_0$. The data are calculated from numerical resolution of the field-dependent free energy equation with  $\Delta$ = 1 and effective masses $m_e$ = 1 and $m_h$ = 2.5 (see text and Ref. \cite{Fo09}). $F_a=7/2$ is the fundamental frequency for this model. (b)  $B/T$
dependence of the first harmonic amplitude $A_a$ for various values of the tunneling probability (hence of the magnetic field). Solid
symbols come from numerical resolution of the semiclassical spectrum, which
fall onto the solid lines corresponding to the Lifshits-Kosevich approximation
given by Eq.~\ref{Eq:ampA1} including contributions of zero area orbits. Dashed
lines correspond to the first order term $l=1$ in Eq.~\ref{Eq:ampA1} or
equivalently to Eq.~\ref{Eq:dHvA}.}
\end{figure}

Contrary to magnetoresistance data, the whole dHvA spectra is quantitatively accounted for
by the Falicov-Stachowiak model in the field range up to 28 T \cite{Au05}. This
latter point is at variance with the data for the linear chain of coupled orbits
considered in Section~\ref{sec:linear_chain}. In order to solve this discrepancy, a
linear chain of compensated orbits, $i.e.$ the 1D network which accounts for the FS in
Fig.~\ref{Fig:SF_networks}(e), has been considered as a first step in Refs.~\cite{Fo09,Fo08}. Indeed, its energy spectrum can be easily deduced from the Pippard's method~\cite{Pi62}.
In this model the effective masses linked to the electron $m_e$ and
hole band $m_h$ are taken to be independent. For simplicity, two
quadratic potentials have been considered to model the FS, i.e. quadratic functions of the
quasi-momenta are assumed for the bands dispersion, as within the free-electron model in two dimensions.
The bottom of the electron band is set at zero energy while the top of the hole
band (inverse quadratic potential) is at $\Delta>0$, with the possibility for
the quasiparticle to tunnel
through the gap between two successive orbits by MB, i. e.
between the pink and green parts of the FS in Fig.~\ref{Fig:SF_networks}(e).
Given an energy $E$, the k-space areas of the closed
electron and hole orbits are respectively given by $S_e=2\pi
m_eE$ and $S_h=2\pi m_h(\Delta-E)$, which are both quantized. The zero field Fermi energy $E_F$ is given by the condition of
compensation $S_e(E_F)=S_h(E_F)$ or $E_F=[m_h/(m_e+m_h)]\Delta$.
The unique fundamental frequency of this system is therefore equal to
$F_a=S_e(E_F)/2\pi=S_h(E_F)/2\pi=m_em_h\Delta/(m_e+m_h)$.

Strikingly, the field-induced chemical potential oscillation, calculated by extremizing the free energy is strongly damped compared to the uncompensated case of Section~\ref{sec:linear_chain} \cite{Fo09,Fo08}. For this reason, Fourier amplitudes can be calculated within the LK formalism. Nevertheless, it is important to notice that, in systems with compensated bands, there is an infinite number of classical orbits contributing to any
given Fourier component since a semi-classical trajectory around a successive electron
and hole pockets has a zero area. Here orbits with zero area can be classified by their
increasing masses $(l-1)m_e+(l-1)m_h$, where $l >$ 1 is an integer.
The amplitude $A_a$ of the Fourier component with frequency $F_a$ is therefore not only dependent
on the orbits composed of one electron or one hole orbit with effective mass $m_e$ or $m_h$, respectively, but also on a series of additional multiple orbits composed of $l-1$ electron and $l-1$ hole orbits plus one electron or one hole orbit, with effective mass
$m_e(l)=lm_e+(l-1)m_h$ or $m_h(l)=(l-1)m_e+lm_h$, respectively. $A_{a}$ can be computed exactly in this model, within the LK theory, and is given by the expression

\begin{eqnarray} \label{Eq:ampA1}
A_{a}&=&\frac{F_a}{\pi}
\Big [
\frac{q_0^2R(m_{e})}{m_{e}}+
\frac{q_0^2R(m_{h})}{m_{h}}
+
\sum_{l\ge
2}
\Big (
\frac{R(m_{e(l)})}{m_e(l)}
+
\frac{R(m_{h(l)})}{m_h(l)}
\Big )
\times
\sum_{n=1}^{2(l-1)}(-1)^np_0^{2n}q_0^{4l-2n-2}\coef(l,n)
\Big ]
\end{eqnarray}

where the combinatorial quantities can be defined by the summations

\begin{eqnarray}\label{Eq:coeff}
\coef(l,n)=\frac{2}{n}\sum_{i=0}^{n/2}\sum_{j=0}^{i}\sum_{k=0}^{n/2-j}
\frac{(-1)^j}{2^{2k}}\times
\bin{n} {2i }
\bin {i}{j}
\bin{n-2j}{2k}
\bin{2k}{k}
\bin{l+k-1}{l-n+j+k}
\bin{l+k-2}{l-n+j+k-1}.
\end{eqnarray}

These positive integers count the number of
non-equivalent orbits for a given mass $m_{e(h)}(l)$
visiting $2n$ pockets on the path. Table 1 displays the numerical values of
$S(l,n)$ computed directly from Eq.~\ref{Eq:coeff}. Damping
factors are given by $R(m_{i(l)}) = R^T_{i,l}R^D_{i,l}R^s_{i,l}$, where $i = e,h$. $p_0$ and $q_0$ are the MB tunneling and reflection probabilities (see Eqs.~\ref{Eq:RT} to~\ref{Eq:Rs}). Even though the contribution of large masses is usually negligible at low field
or high temperature, it can be significant in the large $B/T$ limit
in which case they have to be taken into account, provided the Dingle temperature is small \cite{Fo09}.
Indeed, the temperature dependent factors are close to unity in this limit.

\begin{table}[ht]
\caption{First values of the coefficients $\coef(l,n)$ representing
the number of non-equivalent orbits for a given mass
$m_{e(h)}(l)$ with $2n$ magnetic breakdowns, $1\le n\le 2(l-1)$. }
\centering
\begin{tabular}{| c |  l l l l l l l l l l l l |}
\hline
$_l\backslash^n$ & 1 & 2 & 3 & 4 & 5  & 6 & 7 & 8 & 9 & 10 & 11 & 12
\\
\hline
2 & 2 & 1 &  &  &  & & & & & & &
\\
3 & 2 & 9 & 8 & 1 &  & & & & & & &
\\
4 & 2 & 23 & 68 & 63 & 18 & 1 & & & & & &
\\
5 & 2 & 43 & 264 & 610 & 584 & 228 & 32 & 1 & & & &
\\
6 & 2 & 69 & 720 & 3080 &  6132 & 5930 & 2800 & 600 & 50 & 1 & &
\\
7 & 2 & 101 & 1600 & 10925 & 36980 & 66374 & 64952 & 34550 & 9650 & 1305 & 72 &1
\\
\hline
\end{tabular}
\end{table}

Technically, the summation over all the orbits with zero area is evaluated using
an analogy with an Ising model on a linear chain in magnetic field. Indeed, it
is useful to view each quasiparticle traveling on the
linear chain represented by the FS of Fig.\ref{Fig:SF_networks}(e) as a
particle performing a one dimensional random walk. A set of
positions on the chain $\{x_i\}_{i=0,2n}$ can be defined for a given periodic orbit, with boundary
conditions $x_0=x_{2n}=0$.
These coordinates take integer values (negative and positive) and define the
pocket inside which the quasiparticle is located. In what follows, $x_i$ with
$i$ even (odd) is the position of a quasiparticle in one electron (hole) band. A closed path has an even number of steps $2n$. Given $x_i$,
the particle can also orbit a number $n_i\ge 0$ of times around the surface
before going to the next band by MB. We can rewrite coordinates $x_i$ with mean
of forward/backward variables $\sigma_i=\pm 1$ such as
$x_{i}=x_{i-1}+\sigma_{i}(x_{i-1}-x_{i-2})$. Here $\sigma_i=1$ when
the particle is going forward and $\sigma_i=-1$ when it is going backward. An
adequate set of variables is given by $y_i=x_i-x_{i-1}=\pm 1$ which satisfy
$y_i=\sigma_iy_{i-1}$ or $\sigma_i=y_iy_{i-1}$. All the possible orbits contributing
with a given effective mass can then be counted by summing over all the possible $y_i=\pm$1
restricted to the boundary conditions. It can then be proved that this
combinatorial number is equal to the partition function of an Ising model in a
(complex) magnetic field and which is solvable, leading to Eq.~(\ref{Eq:coeff}).
In two-dimensions, compensated networks such as the FS reported in Fig.~\ref{Fig:SF_networks}(d)
could also lead to non-negligible contribution of zero-area orbits, but their
explicit evaluation is difficult since a similar analogy with the previous calculation would lead
equivalently to the computation of the partition function for a two-dimensional Ising model in magnetic field.

The relevance of Eq.~\ref{Eq:ampA1} is evidenced in Fig.~\ref{Fig:TFcomp}. Oscillations data in this figure are calculated from numerical resolution of the field-dependent free energy equation given in Ref.~\cite{Fo09}. As expected, Fourier analysis  exhibits only one frequency $F_a$ and harmonics. Fig.~\ref{Fig:h1} compares the temperature dependence of $A_a$ deduced from numerical resolution (solid symbols) to the predictions of Eq.~\ref{Eq:dHvA}, which neglects any contribution of orbits with $l > $ 1 (dashed lines). A growing discrepancy is observed as the temperature decreases, except in the absence of MB ($q_0$ = 1) in which case only the basic closed orbits contributes. In contrast, an excellent agreement with Eq.~\ref{Eq:ampA1} (solid lines) is observed. Shortly speaking, at variance with uncompensated orbits networks, the LK formula accounts for the quantum oscillations in a linear chain of compensated orbits (one-dimensional network), provided the contribution of all MB orbits is taken into account for each frequency.

\section{Conclusion}

Although few puzzling results are observed, as reported in Section~\ref{sec:puzzling}, numerous organic conductors achieve Fermi surfaces that can be regarded as model systems for the study of quantum oscillations in networks of orbits coupled by magnetic breakdown. Two types of networks can be distinguished, namely the linear chain of coupled orbits and the networks of compensated orbits (see few examples in Fig.~\ref{Fig:SF_networks}). These networks are considered in Sections~\ref{sec:linear_chain} and~\ref{sec:compensated}, respectively.

A user friendly analytical tool, taking into account the chemical potential oscillations in magnetic field, has recently been proposed to account for dHvA oscillation spectra of the linear chain of coupled orbits \cite{Au12}. The main feature of the derived formulae is, besides standard Lifshits-Kosevich terms, the presence of second order corrections liable to account for the observed Fourier components that are 'forbidden' in the framework of the semiclassical Falicov-Stachowiak model. This model has been successfully implemented in the dHvA oscillations data of $\theta$-(ET)$_4$CoBr$_4$(C$_6$H$_4$Cl$_2$). Obviously, this encouraging result needs to be confirmed in the case of other relevant compounds. In addition, even stronger deviations are observed for SdH spectra which are still unexplained.

Regarding compensated orbits, either one- (see Fig.~\ref{Fig:SF_networks}(e)) or two-dimensional (see Fig.~\ref{Fig:SF_networks}(d), (f)) networks can be observed. At variance with the above case, oscillations of the chemical potential are strongly damped in one-dimensional networks and the Lifshits-Kosevich theory applies in this case, provided multiple orbits are taken into account. Unfortunately, to our knowledge, no experimental data relevant to one-dimensional networks are available to check this result. Nevertheless, dHvA oscillations spectra of the two-dimensional network achieved by the (ET)$_8$[Hg$_4$Cl$_{12}$(C$_6$H$_5$Br)$_2$] compound are nicely accounted for by the Falicov-Stachowiak model in magnetic fields below 28 T which is in line with the above reported damping of the chemical potential oscillations. However, SdH oscillations evidences strong deviations from the Lifshits-Kosevich model, including the presence of 'forbidden' Fourier components with high amplitude, besides the contribution of quantum interference. These results still require interpretation.

Regarding new problematics, promising perspective are brought about by organic metals with two different donors planes, few examples of which can be found in the literature. In few cases, a metal-insulator transition is observed as the temperature decreases \cite{Ma07,Ak11} suggesting two insulating planes at low temperature. Compounds with one metallic and one insulating layer with either compensated orbits \cite{Zo11} or, as reported in Section~\ref{sec:linear_chain}, linear chain of coupled orbits \cite{Au12,Sh11} have also been reported. In these cases, a strongly two-dimensional behaviour is observed. More appealing case is provided by unit cell with two different metallic layers \cite{Ly01,Vi07b,Zh09}, hence with two different Fermi surfaces \cite{Vi07b}. Such a structure could be relevant to the bilayer splitting phenomenon reported for cuprate superconductors \cite{Po08} and addresses the question of a three-dimensional Fermi surface definition. To our knowledge, quantum oscillations data at high field have only been reported for $\theta$-(BEDT-TSF)$_4$HgBr$_4$(C$_6$H$_5$Cl) for which complicated spectra, with no clear frequency combinations are observed \cite{Vi07b}. In any case, the oscillatory data cannot be interpreted on the basis of a mere addition of the contributions of the two Fermi surfaces pertaining to each of the two donors planes. Synthesis of other compounds and further experiments are needed to get a better insight on this issue.




\section*{Acknowledgements}
We wish to warmly acknowledge David Vignolles, Fabienne Duc and Lo\"{\i}c Drigo
of the LNCMI, Toulouse; Vladimir N. Laukhin and Enric Canadell of the
ICMAB, Barcelona; Rustem B. Lyubovskii, Rimma N. Lyubovskaya and Eduard B.
Yagubskii of the IPCP, Chernogolovka; Tim Ziman of the Institut Laue-Langevin, Grenoble, and all our collaborators whose names appear in Refs.
\cite{Au12,Vi08,Fo98b,Vi09,La11,Vi06,Au04,Au06,Vi10,Vi07,Fo98a,Fo05,Pr02,Vi03,Au05,Au11,Vi07b}. Part of this work has been supported by EuroMagNET
II under the EU
contract number 228043.

\end{document}

